\documentstyle[12pt]{article}
\topmargin - 1.5cm
\leftmargin - 1cm
\begin{document}

\title{ Resonance absolute quantum reflection at selected energies }
\author{V. M. Chabanov and B. N. Zakhariev}
\date{27 March 2001}
\maketitle
\begin{center}
Laboratory of Theoretical Physics, \\
Joint Institute for Nuclear Research,  \\ 
141980 Dubna, Russia   \\
email: zakharev@thsun1.jinr.ru
\end{center}

\begin{abstract}
The possibility of the resonance reflection (100 \% at maximum) is 
revealed.  The corresponding exactly solvable models 
with the controllable numbers of resonances, their positions and widths  
are presented.                                                
\end{abstract}

The foundation of quantum mechanics is, after all, 
the laws of wave motion in different potentials. The most interesting are 
the potentials with the especial qualitative properties because they  give
a deeper insight into the peculiarities of the microworld.  We have 
revealed the 
phenomenon of {\it total reflection} at selected energy points by specific 
potentials.  It is remarkable that this occurs even for {\it 
above-barrier} motion.  The possibility itself of such a resonance 
reflection has surprisingly never been  mentioned since the very 
beginning of the wave mechanics.  

Let us consider the one-dimen\-sional 
 Schr\"odinger equation on the whole axis ($-\infty <  x < \infty $) 
with time-independent potentials.  
For our purpose, we shall need the potentials of Neumann-Wigner type which 
have bound states embedded into continuum (BSEC)  on the 
{\it semi-axis} ${0 \le x<\infty }$  \cite{Mos,IP,III,ChZ}. 
The expression for the potential in the simplest case of the only bound 
state is the following  

 \begin{eqnarray}
V(x)=-2 \frac{d}{dx}\{\frac{c^2 \sin(k x)^{2}}{k^2 [1 + \frac{c^2}{k^2}
(\frac{x}{2} - \frac{\sin(2 k x)}{4 k})]} \}, \nonumber \\
k = \sqrt{E_{BSEC}}.
\label{pot}
\end{eqnarray}
The BSEC wave function normalized to unity at the positive energy value 
$E=E_{BSEC}>0$ has the form 
\begin{eqnarray} 
\Psi(x,k) = 
\frac{c \sin(k x)}{k [1 + \frac{c^2}{k^2} (\frac{x}{2} - \frac{\sin(2 k 
x)}{4 k})]},
\label{wf} 
\end{eqnarray} 
see the right-hand-side of Fig.1 for $x \ge 0$.
The parameter $c$ is the derivative of the BSEC function (\ref{wf}) at the 
origin.  Both the functions, $V(x) $ (\ref{pot}) and $\Psi(x,k)$ 
(\ref{wf}), decrease asymtotically $\sim \frac{1}{x}$, as $x \rightarrow 
\infty$. Every antinode of the BSEC corresponds to well-barrier block of 
the potential confining the wave from the right \cite{CZS}.   Besides BSEC, there is another 
linear independent solution at $E=E_{BSEC}$ which diverges asymptotically 
as $x \rightarrow \infty$.  This solution is unphysical and must be 
omitted.

If continued to the {\it whole axis} with $V(x<0)\equiv 0$,
these potentials (\ref{pot}) lose their confinement property  
because the waves now leak out to the left semi-axis where they move 
freely.  Instead this potentials acquire the new remarkable 
feature : they give a total reflection at the BSEC energies for 
the waves incident from the left. 

This can be explained in the following way.  
The BSEC-potentials  confine wave on the semi-axis 
 ${0 \le x<\infty }$ at the chosen energy $E=E_{BSEC}$.  They forbid the 
wave propagation to the right.  So, it is natural to expect complete 
reflection of the waves incident on these potentials from the left. To be 
more precise, the only physically acceptable solution on the whole axis  
must coincide with the BSEC (\ref{wf}) on the semi-axis $x \ge 0$ which 
asymptotically decreases.  Its smooth continuation to the negative 
semi-axis is a free wave $c \sin(k x)/k$ as  shown in Fig.1. This sine is 
a combination of incident and outgoing waves: $c [\exp(i k x) - \exp(- i k 
x)]/2ik$.  The normalized solution has a unit amplitude of incident 
wave and the reflected wave  with the reflection coefficient 
$|R(E)_{BSEC}|=1$. 

At the energies differing from $E=E_{BSEC}$ (pinned 
out point) the potentials (\ref{pot}) do not confine the waves on half-axis 
and cannot be totally reflective.   The modulus of reflection coefficient 
$|R(E)|$ is shown in Fig.2. Pay also attention  to  the 
dependence of $|R(E)|$ on the parameter $c$. A physical
sense of this parameter is that it determines the measure of localization 
of the BSEC near the origin. With increase in $c$, the BSEC 
function (\ref{wf}) is concentrated at $x=0$ and becomes 
$delta$-function in the limiting case. Large $c$ values correspond to a 
width  of  the reflection resonance. On the contrary, for the 
smaller $c$  the resonance peak occurs narrower. Thus, one can control 
the width of the resonance peak at $E=E_{BSEC}$ by varying the parameter 
$c$.  We can also control the number of the resonance points including 
their positions.  Indeed, they are associated with the BSEC energies on the 
half-axis which can be created by using the inverse problem formalism 
(exact solutions generalizing  Eqs.  (\ref{pot}, \ref{wf}).  So we get a 
new class of potentials corresponding to the exactly solvable models with 
resonance reflection.

It is worth to mention that tails of BSEC-potentials on $a \le x \ge 
\infty$ for any $a$  are responsible for the wave confinement. 
So, there can be added almost arbitrary potentials at the left side which 
would not change the position of the reflection resonance, but deform its 
shape.  Particularly, one can expect a tunneling resonance (with 100\% 
penetrability) near the non-penetrability point. But this details will be 
considered elsewhere.

There is possible the additional control of BSEC. The space-localization 
of wave accumulation can be shifted by different combinations of 
abovementioned potential blocks : well-barrier block shifts maximum of the 
corresponding wave bump to the left ($<-$) and barrier-well pushes one
to the right ($->$) \cite{CZS}. It is  important that wave function knots 
at $E=E_{BSEC}$ coincide with even knots of the potential. 
The change of wave derivatives at
the knots for each bump and smooth connections of the neighbor 
wave-bumps result in decreasing (see Fig.1) (or increasing in the case $->$)
 in their relative amplitudes.  So, for example, the maximum of BSEC's 
bump amplitudes can be shifted to the right substituting  some 
($<-$)-blocks by ($->$)-blocks:  $$ (->...->max|ampl.  
\Psi_{BSEC}|<-...<-..)$$

Of course, the periodic potentials on the half-axis also have the property 
of total reflection. This occures at the energy values belonging to the 
 forbidden spectral zones of  the same periodic potential continued to the 
whole axis, which seems to be almost evident. However periodic 
potentials are not quadratically integrable as potentials with BSEC and
have whole bands of total reflection, unlike our case when absolute
non-penetrability occures at the chosen energy points.

 One should not  confuse the phenomenon with 
the total reflection of the waves incident at some angle to a plane of 
demarcation of different optical mediums.  In our case, the total 
reflection is even for incident waves  perpendicular to this plane.  The 
same also concerns the Bragg reflection from parallel 
crystal lattice planes.  

The resonance reflection is also possible in multichannel case on the whole 
axis: e.g., for M coupled Schr\"odinger equations \cite{Ann}. Here, the 
phenomenon appears to be even more diversified. Unlike the
one-channel case there is possibility for short range (exponentially 
decreasing) BSEC potential matrices. There can be different kinds of 
BSEC states on the half-axis at the fixed energy value.  In the case of M 
degenerated BSEC states there will be resonance  100\% reflection for any 
combination of incident waves in different channels. There can also coexist 
M-m BSEC and m scattering states at $E=E_{BSEC}$ on the half-axis. Then on 
the whole axis there will be m linearly independent combinations of 
incident waves with total reflection at $E_{BSEC}$ and M-m combinations 
with the comparatively weak  reflection. More details will be published 
elsewhere.

 One might be somewhat surprised that the 
resonance reflection  was not known before in spite of long and intensive 
investigations of quantum scattering. Maybe it was because of the fact
 that the possibility itself of this phenomenon is  due to existence of 
potentials with BSEC which were better understood in the 
inverse problem approach. Particularly, striking was the property of wave 
confinement with positive energy {\it above the potential barriers}, as in 
Fig.1.  Besides it was necessary to transform BSEC potentials so 
that they lose the property to keep BSEC, to admit the wave leakage to the 
left. Only at the expense of this transformation, the potential 
acquires the new ability of absolute selective reflectivity.

  The classes of exact solvable models considered above contribute 
significantly to our theoretical notions in wave dynamics and interference 
for quantum design \cite{IP,Les,ZS},  optics, acoustics, radiowave 
propagation and possible applications.

In conclusion we want to mention an interesting  paper \cite{ZA} about 
special potentials with reflection resonances, but with $|R(E)|<1$ at
maximum.

\newpage

Figure Captions

FIG.1. The only acceptable solution at $E=E_{BSEC}=1; \enskip c=1$ (solid 
line) and the corresponding  potential which is zero at the negative 
semi-axis and equal to BSEC-potential at the positive semi-axis. Pay 
attention to the coincidence of  BSEC knots with the even knots of 
BSEC-potential.  It is this exact correlation that provides the confinement 
the waves from the right at the pinned out point $E_{BSEC}$.

FIG.2. The modulus of the reflection coefficient $|R(E)|$ for potentials
having bound states embedded into continuum spectrum (BSEC) on the
half-axis $0\le x < \infty $  at energy $E_{bound}=10$ with resonance
nonpenetrability at this point for waves on the whole axis. With increasing
of BSEC spectral weight parameter  $C$ the width of resonance in $|R(E)|$
near the $E_{bound}=10$ becomes greater. For   $C\rightarrow 0$ this
width converges to zero: the dashed line corresponds to the limiting
peak in $R(E)$.

\end{document}